\documentclass[aps,prb,twocolumn,showpacs,superscriptaddress,floatfix,longbibliography]{revtex4-1}

\newcommand{\ket}[1]{\ensuremath{|#1\rangle}}
\newcommand{\bra}[1]{\ensuremath{\langle#1|}}
\newcommand{\angstrom}{\text{\normalfont\AA}}

\usepackage{graphicx}
\usepackage{dcolumn}
\usepackage{bm}
\usepackage{amssymb}
\usepackage[version=4]{mhchem}
\newcolumntype{C}[1]{>{\centering\arraybackslash}p{#1}}
\usepackage{xfrac}
\usepackage{gensymb}
\usepackage{xcolor}
\usepackage{color}
\usepackage{multirow}
\usepackage{comment}
\usepackage{times}
\usepackage[varg]{txfonts}
\usepackage{mathrsfs}
\usepackage{amsmath,amssymb,bm,mathtools}
\usepackage{amsfonts}
\usepackage{booktabs}
\usepackage{natbib,hyperref}
\usepackage{scrextend}
\usepackage{dirtytalk}
\usepackage{soul}
\usepackage{epstopdf}
\usepackage{color}
\usepackage{array}
\usepackage{multirow}
\usepackage{tcolorbox}
\usepackage[percent]{overpic}
\usepackage{bbm}
\usepackage{widetable}
\usepackage{array}
\usepackage{xcolor,colortbl}
\usepackage{lipsum}

\definecolor{cred}{RGB}{188,55,84}
\hypersetup{
	colorlinks=true, 
	linkcolor=blue,  
	citecolor=cred,
	urlcolor=cred,
}

\begin{document}
\title{Spin-orbital order and excitons in magnetoresistive HoBi}

\author{J.~Gaudet}
    \email[Correspondence email address: ]{Jonathan.Gaudet@nist.gov}
    \affiliation{Institute for Quantum Matter and Department of Physics and Astronomy, Johns Hopkins University, Baltimore, MD 21218, USA}
    \affiliation{Center for Neutron Research, National Institute of Standards and Technology, MS 6100 Gaithersburg, Maryland 20899, USA}
    \affiliation{Department of Materials Science and Eng., University of Maryland, College Park, MD 20742-2115}

\author{H.-Y.~Yang}
    \affiliation{Department of Physics, Boston College, Chestnut Hill, Massachusetts 02467, USA}

\author{E.~M.~Smith}
\affiliation{Department of Physics and Astronomy, McMaster University, Hamilton, ON L8S 4M1, Canada}

\author{T.~Halloran}
    \affiliation{Institute for Quantum Matter and Department of Physics and Astronomy, Johns Hopkins University, Baltimore, MD 21218, USA}
    
\author{J.~P.~Clancy}
\affiliation{Department of Physics and Astronomy, McMaster University, Hamilton, ON L8S 4M1, Canada}

\author{J.~A.~Rodriguez-Rivera}
\affiliation{Center for Neutron Research, National Institute of Standards and Technology, MS 6100 Gaithersburg, Maryland 20899, USA}
\affiliation{Department of Materials Science and Eng., University of Maryland, College Park, MD 20742-2115}
    
\author{Guangyong Xu}
\affiliation{Center for Neutron Research, National Institute of Standards and Technology, MS 6100 Gaithersburg, Maryland 20899, USA}

\author{Y.~Zhao}
\affiliation{Center for Neutron Research, National Institute of Standards and Technology, MS 6100 Gaithersburg, Maryland 20899, USA}
\affiliation{Department of Materials Science and Eng., University of Maryland, College Park, MD 20742-2115}

\author{W.~C.~Chen}
\affiliation{Center for Neutron Research, National Institute of Standards and Technology, MS 6100 Gaithersburg, Maryland 20899, USA}

\author{G.~Sala}
\affiliation{Spallation Neutron Source, Second Target Station, Oak Ridge National Laboratory, Oak Ridge, TN, 37831, USA}

\author{A.~A.~Aczel}
\affiliation{Neutron Scattering Division, Oak Ridge National Laboratory, Oak Ridge, Tennessee 37831, USA}

\author{B.~D.~Gaulin}
\affiliation{Department of Physics and Astronomy, McMaster University, Hamilton, ON L8S 4M1, Canada}
\affiliation{Canadian Institute for Advanced Research, 661 University Avenue, Toronto, Ontario M5G 1M1, Canada.}
\affiliation{Brockhouse Institute for Materials Research, Hamilton, ON L8S 4M1 Canada}

\author{F.~Tafti}
    \affiliation{Department of Physics, Boston College, Chestnut Hill, Massachusetts 02467, USA}
    
\author{C.~Broholm}
\affiliation{Institute for Quantum Matter and Department of Physics and Astronomy, Johns Hopkins University, Baltimore, MD 21218, USA}
\affiliation{Center for Neutron Research, National Institute of Standards and Technology, MS 6100 Gaithersburg, Maryland 20899, USA}
\affiliation{Neutron Scattering Division, Oak Ridge National Laboratory, Oak Ridge, Tennessee 37831, USA}

\date{\today} 

\begin{abstract} 
The magnetism of the rock-salt $fcc$ rare-earth monopnictide HoBi, a candidate topological material with extreme magnetoresistance, is investigated. From the Ho$^{3+}$ non-Kramers $J$=8 spin-orbital multiplet, the cubic crystal electric field yields six nearly degenerate low-energy levels. These constitute an anisotropic magnetic moment with a Jahn-Teller-like coupling to the lattice. In the cubic phase for $T>T_N~=~5.72(1)~K$, the paramagnetic neutron scattering is centered at $\mathbf{k}=(\frac{1}{2}\frac{1}{2}\frac{1}{2})$ and was fit to dominant antiferromagnetic interactions between Ho spins separated by $\{100\}$ and ferromagnetic interactions between spins displaced by $\{\frac{1}{2}\frac{1}{2}0\}$. For $T<T_N$, a type-II AFM long-range order with  $\mathbf{k}=(\frac{1}{2}\frac{1}{2}\frac{1}{2})$ develops along with a tetragonal lattice distortion. While neutron diffraction from a multi-domain sample cannot unambiguously determine the spin orientation within a domain, the bulk magnetization, structural distortion, and our measurements of the magnetic excitations all show the easy axis coincides with the tetragonal axis. The weakly dispersive excitons for $T<T_N$ can be accounted for by a spin Hamiltonian that includes the crystal electric field and exchange interactions within the Random Phase Approximation. 
\end{abstract}

\maketitle

\section{introduction}

\begin{figure}[!ht]
\linespread{1}
\par
\includegraphics[width=2.7in]{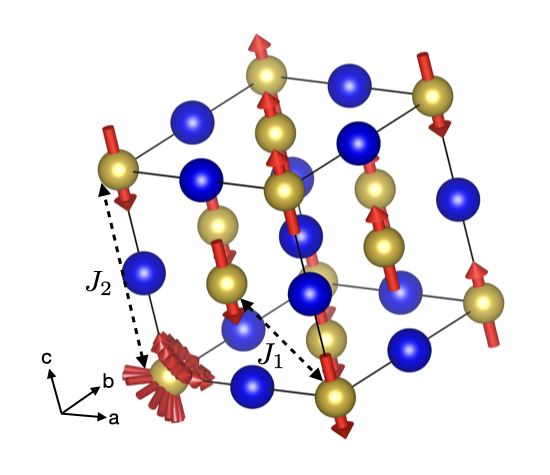}
\par
\caption{The rock-salt structure of the rare-earth monopnictide HoBi. Yellow and blue spheres respectively correspond to Ho and Bi. Spins interacting through the $J_1$ and $J_2$ exchange interaction are shown by the dashed black arrows. The $\mathbf{k}=(\frac{1}{2}\frac{1}{2}\frac{1}{2})$ magnetic order of the Ho$^{3+}$ spins is represented by the red arrows. The local spin orientations of the Ho$^{3+}$ spins that are consistent with neutron diffraction are indicated for the Ho ion at (0,0,0). The magnetization, structural distortion, and inelastic neutron scattering however, provide clear evidence for easy [001] axis anisotropy.}
\label{NucStruc} 
\end{figure}

In spite of their structural simplicity, the $fcc$ rare-earth monopnictides (see Fig.~\ref{NucStruc}), RX (R=Ce to Yb and X=N, As, P, Sb, and Bi~\cite{Duan2007,Petit2010}), display a wide variety of anisotropic magnetism and electronic transport properties. The lattice parameter varies by 30$\%$ across the pnictide series and this provides opportunities to tune the relative strength of crystal field and exchange interactions. In the 1960s to 1980s, the rare-earth monopnictides were studied to understand magnetic phases driven by oscillatory and highly anisotropic Ruderman-Kittel-Kasuya-Yosida (RKKY) exchange interactions ~\cite{Child1963,Turberfield1971,Birgeneau1973,schobinger1974magnetic,Fischer_1977,Heer_1979}. Work on CeSb for example gave rise to an extensive literature on the anisotropic nearest and next nearest neighbor Ising model (ANNNI)\cite{SELKE1988213}. This  work also resulted in progress towards a quantitative understanding of their anisotropic exchange interactions\cite{PhysRevB.50.965}.

A recent resurgence of interest in these rare-earth monopnictides is driven by their extreme magnetoresistance (XMR) and resistivity plateaus, and the possible connection to the 3D topological state of the non-magnetic lanthanum monopnictides LaX~\cite{tafti2016resistivity,tafti2016temperature}. LaAs, LaSb, and LaBi have unsaturated XMR arising from near perfect electron-hole compensation and there is a topological transition from a trivial electronic band structure in LaAs to a topologically non-trivial band structure in LaBi~\cite{guo2016charge,yang2017extreme,lou2017evidence,oinuma2017three,Nummy2018}. Several studies have confirmed the presence of protected surface states in LaBi~\cite{niu2016presence,singha2017fermi,Nayak2017,feng2018experimental}. Since then, extensive works have been devoted to characterizing the XMR and topological states of various RX including for example CeX, HoX, and PrX. XMR has been found in each reported magnetic RX with characteristics that depend on the rare-earth ion~\cite{pavlosiuk2018magnetoresistance,liang2018extreme,wu2017large,ye2018extreme,wang2018extremely,wang2018unusual,lyu2019magnetization,wu2019multiple,hosen2020observation}. The stabilization of topological non-trivial electronic bands generating protected surface states was proposed for several of the magnetic monopnictides~\cite{Neupane_2016,guo2017possible,wu2019anomalous,wu2019multiple}. 

Here we study the magnetism of HoBi using modern neutron scattering techniques to gain insights into its unique magneto-transport properties~\cite{wu2019multiple,yang2018interplay}. Consistent with previous works~\cite{hulliger1984low,FISCHER1985551,fente2013low}, we confirm the antiferromagnetic (AFM) $\mathbf{k}~=~(\frac{1}{2}\frac{1}{2}\frac{1}{2})$ structure and the associated tetragonal lattice distortion. Due to multi-domain averaging, our single-crystal neutron diffraction cannot unambiguously determine the local spin anisotropy of the $\mathbf{k}~=~(\frac{1}{2}\frac{1}{2}\frac{1}{2})$ AFM structure. However, we could resolve this ambiguity by measuring and modeling the magnetic excitations of HoBi, which take the form of weakly propagating spin-orbital excitons whose energies and intensities are sensitive to the local orientation of the Ho$^{3+}$ moments. Using this method, we found the $\mathbf{k}~=~(\frac{1}{2}\frac{1}{2}\frac{1}{2})$ AFM structure has an Ising local spin anisotropy, which is consistent with the Ising easy-axis bulk magnetization and the tetragonal distortion. Through analysis of the paramagnetic diffuse scattering of HoBi and the crystal field excitons in the low $T$ ordered state, we obtain a spin Hamiltonian with comparable crystal field (CEF) and exchange energy scales.

\section{Experimental Methods}
HoBi single crystals with mass of 10-50~mg were grown following a previously published procedure~\cite{yang2018interplay}. Single crystal low-temperature X-ray diffraction was performed using a Huber four-circle diffractometer with a Rigaku Rotaflex 18~kW rotating copper anode X-ray generator and a Bicron point detector. We used a Ge (111) monochromator with d$_{111}$~=~3.266~\angstrom. The sample was aligned for diffraction in the (HHL) plane and mounted in a closed cycle cryogenic system with a base temperature of 2.17~K.  

We performed thermal neutron diffraction using the HB-1A triple-axis instrument at Oak Ridge National Laboratory. We used PG filtered 14.5~meV neutrons, and collected rocking scans at all accessible magnetic and nuclear Bragg positions in the (HHL) plane. Polarized neutron diffraction measurements were conducted with the triple-axis instrument BT-7 at the Center for Neutron Research (NCNR), NIST. Nuclear spin-polarized $^3\ce{He}$ gas was used to polarize the incident neutron beam and to analyze the polarization of scattered neutrons~\cite{chen2009applications,chen2014}. Horizontal guide fields were present throughout the beam path to allow measurements of the spin-flip (SF) and non-spin-flip (NSF) scattering cross-sections for incident neutron spins polarized parallel to momentum transfer ${\bf Q}$. The flipping ratio measured at nuclear Bragg peaks was greater than 30. 

Cold neutron triple-axis experiments were performed using the SPINS and the MACS spectrometers at the NCNR. On both instruments we employed a fixed final neutron energy $E_f=3.7$~meV or 5~meV and measured the elastic and inelastic scattering for a single crystal of HoBi aligned for scattering within the (HHL) and the (HK0) plane in two different experiments. For the $E_f=3.7$~meV configuration, we used polycrystalline cooled Be and BeO filters before and after the sample, respectively. For the 5~meV configuration we only used a Be filter after the sample while the incident beam from the cold neutron source was unfiltered. For both experiments, we co-mounted 11 HoBi single crystals on an aluminum mount. We acquired background data using an identical mount without HoBi crystals. We used an "orange" $4$He flow cryostat to reach a base temperature of 1.6~K for these experiments. 

For the highest energy resolution and energy transfer, we performed time-of-flight neutron scattering experiments using the CNCS spectrometer at Oak Ridge National Laboratory. There we co-aligned two HoBi single crystals on an aluminum mount and collected inelastic neutron scattering data with fixed incident energy $E_i= 25$~meV at $T~=~13~K$ with a total proton charge of 47 C. We used the high flux mode of operation of CNCS with a Fermi Chopper, Chopper 2, Chopper 3, and a Double Disk frequency of 60, 60, 60, 300, and 300~Hz respectively. The energy resolution (FWHM) at the elastic line for this configuration is 2.0(1)~meV. Finally, we note that the error bars associated with the neutron scattering experiments represent one standard deviation.

Both the magnetization and heat capacity measurements presented here were performed in a Quantum Design physical properties measurement system (PPMS). We used a PPMS dilution refrigerator option for the low-temperature heat capacity.

\section{Results and Analysis}
\subsection{1st order phase transition}

\begin{figure}[!ht]
\linespread{1}
\par
\includegraphics[width=3.4in]{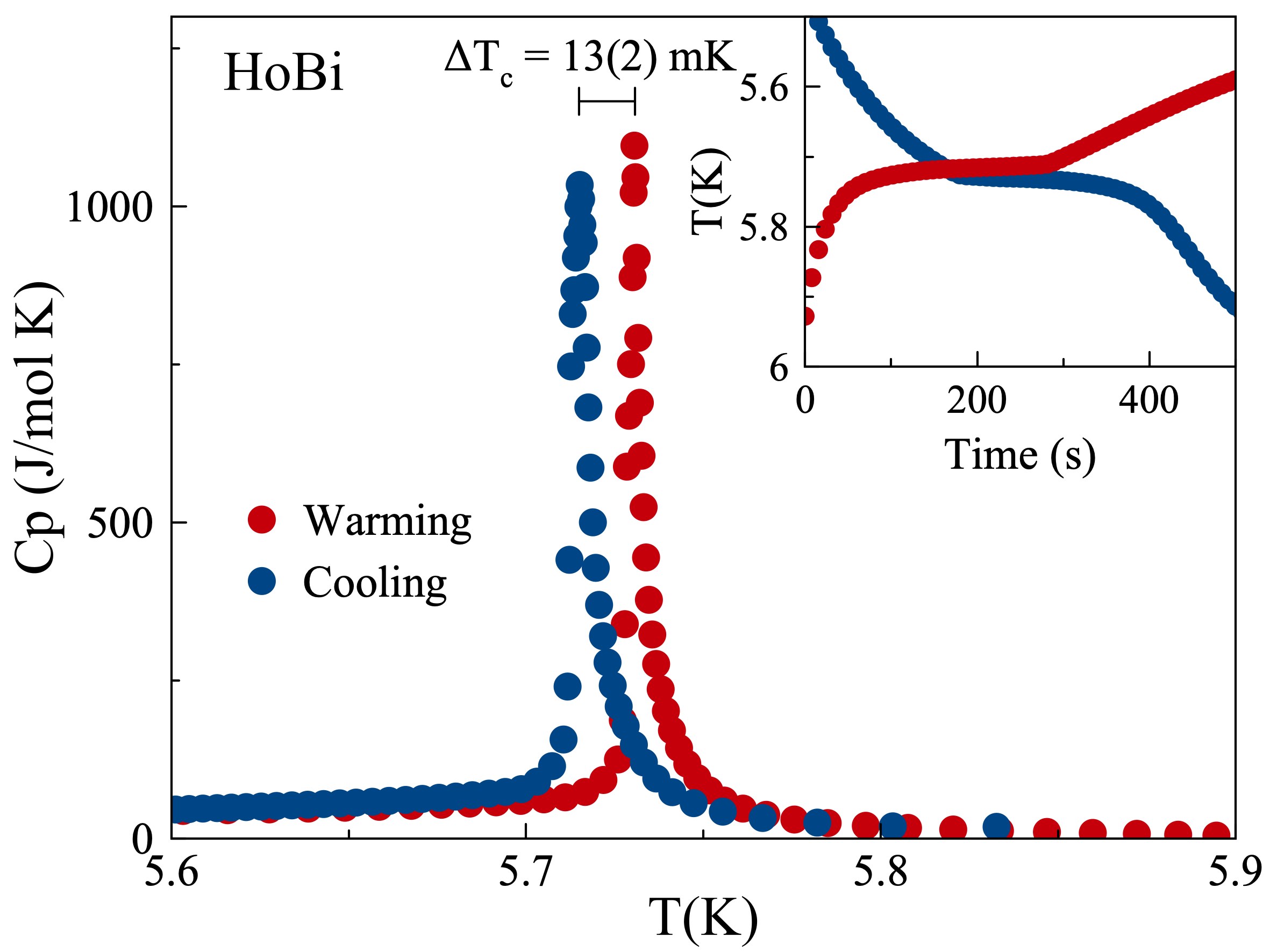}
\par
\caption{Low temperature heat capacity of HoBi collected using the long-pulse method. The red and blue curves respectively correspond to the warming and cooling protocol and shows a thermal hysteresis of 13(2)~mK. The observation of a plateau at $T_N$ in the heating profile for both warming and cooling protocol (top inset panel) suggests a 1st order phase transition in HoBi.}
\label{CpHoBi}
\end{figure}

The thermodynamic properties of HoBi were previously reported and a long-range $\mathbf{k}=(\frac{1}{2}\frac{1}{2}\frac{1}{2})$ antiferromagnetic (AFM) order is known to occur concomitantly with a structural distortion around $T_N~=~5.7~K$~\cite{hulliger1984low,FISCHER1985551,yang2017extreme}. The order of the transition, however, remains unknown. To determine the order of the phase transition, we measured the temperature dependent specific heat capacity using the long-pulse heat method~\cite{ScheieLongPulse}. The resulting $C_p$ data for HoBi is reported in Fig.~\ref{CpHoBi} for both warming and cooling protocols. A sharp peak with a thermal hysteresis of 13(2)~mK is observed in $C_p$. Correspondingly the inset shows a distinct plateau in the temperature versus time curves during heating and cooling. These observations indicate a 1st order phase transition at $T_N$ in HoBi.\\

\subsection{Paramagnetic phase}
\label{diffusetext}
\begin{figure}[tbp]
\linespread{1}
\par
\includegraphics[width=3.4in]{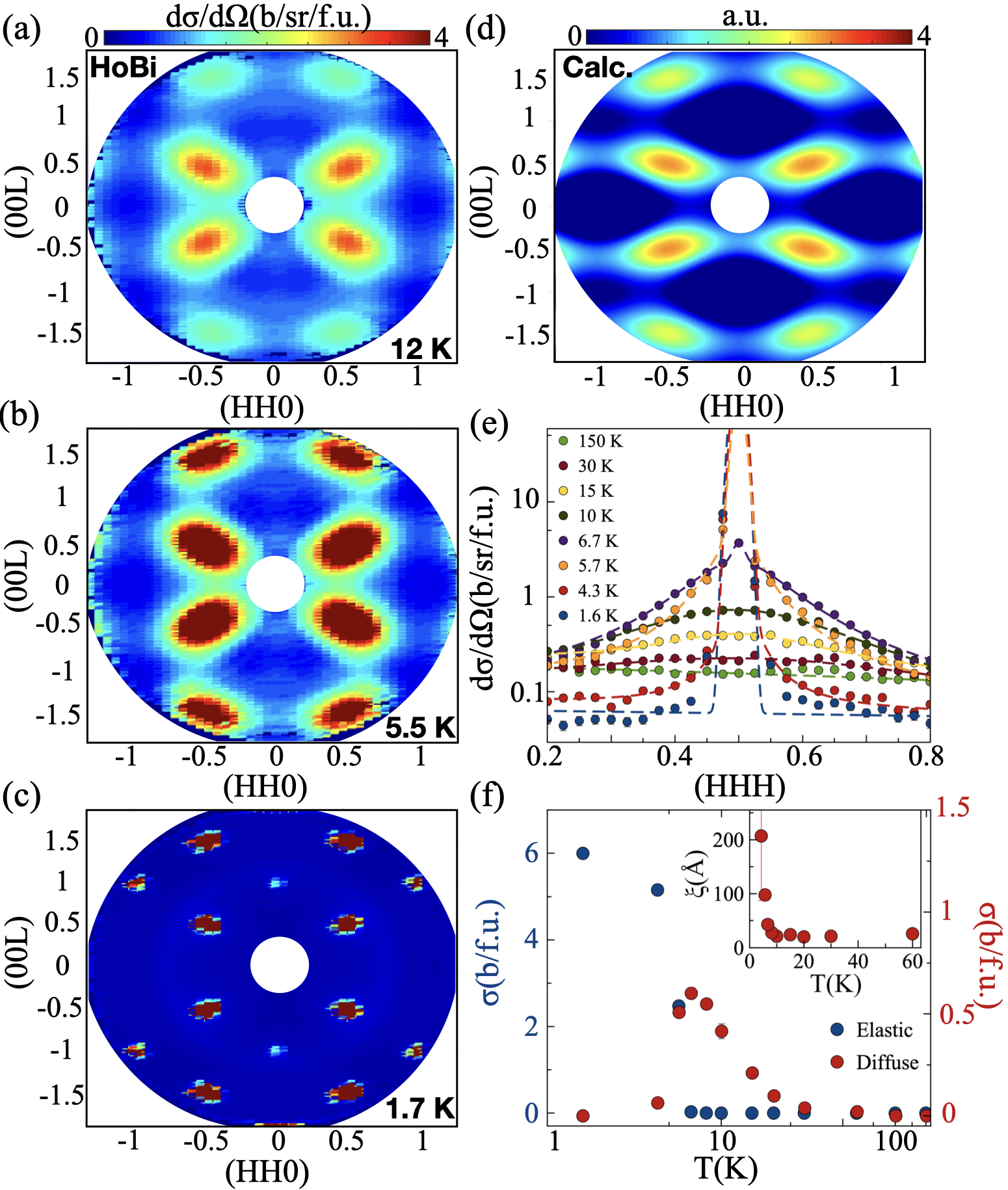}
\par
\caption{The elastic diffuse neutron scattering from HoBi measured in the (HHL) reciprocal lattice plane at (a) 12~K, (b) 5.5~K, and (c) 1.7~K with an incident neutrons energy of 3.7~meV . The scattering for panels (a,b,c) have been symmetrized to increase statistics. (d) Calculated paramagnetic diffuse scattering with J$_1$/J$_2$~=~-2.17 on an $fcc$ lattice where J$_1$ is the first n.n. ferromagnetic interaction and J$_2$ is the 2nd n.n. antiferromagnetic interaction. Panel (e) is the elastic neutron scattering near the $\mathbf{Q}~=~(\frac{1}{2}\frac{1}{2}\frac{1}{2})$ Bragg peak acquired through scans along the the (HHH) direction. The data in panel (e) were fitted using a Lorentzian function for the diffuse scattering and a Gaussian function for the resolution limited Bragg component. The inferred integrated intensity for each component of the scattering are plotted in panel (f) as a function of temperature. The temperature dependence of the magnetic correlation length is plotted in the inset panel of (f).}
\label{Diffuse} 
\end{figure}

To determine the magnetic interactions leading to this phase transition, we mapped the neutron elastic scattering for momentum transfer $\mathbf{Q}$ covering the (HHL) plane and for temperatures between 150~K and 1.6~K. Representative data sets are shown in Fig.~\ref{Diffuse}.

In the cubic paramagnetic phase for $T~=~12{~\rm K}~>~T_N=~5.72(1)~K$, the scattering is broad in $\mathbf{Q}$ and is centered at $\mathbf{k}=(\frac{1}{2}\frac{1}{2}\frac{1}{2})$ positions ((Fig.~\ref{Diffuse}(a)). This indicates short-range AFM correlations preceding the long-range order. The "butterfly" pattern of paramagnetic diffuse scattering is consistent with the equal time structure factor ${\cal S}({\bf Q})$ of an $fcc$ Heisenberg paramagnet with FM interactions between the first nearest-neighbor (n.n.) Ho$^{3+}$ ions ($J_1$), and AFM interactions between the 2nd n.n. ($J_2$). Dashed lines in Fig.~\ref{NucStruc} indicate the lattice geometry associated with these interactions. The scattered intensity was modeled using ${\cal I}({\mathbf  Q})=\frac{2}{3}N|f({\mathbf Q})|^2\sum_{ij}\langle{\mathbf S}_i\cdot {\mathbf S}_j\rangle \cos( {\mathbf Q} \cdot {\mathbf r}_{ij})$ where $N$ is the number of spins, ${\bf r}_{ij}$ is the displacement vector from Ho$^{3+}$ site $j$ to $i$, and $f(\mathbf{Q})$ is the Ho$^{3+}$ atomic form factor~\cite{Blume1962}. Including only self-correlations and correlations between spins separated by $\{100\}$ and $\{\frac{1}{2}\frac{1}{2}0\}$, a ratio of $\langle {\bf S}_i\cdot{\bf S}_j\rangle_{\{100\}}/\langle {\bf S}_i\cdot{\bf S}_j\rangle_{\{\frac{1}{2}\frac{1}{2}0\}}=-2.2(2)$ was obtained at $T~=~12~K$. The calculated magnetic diffuse scattering corresponding to the best fit shown in Fig.~\ref{Diffuse}(d) accounts for all major features in the data (Fig.~\ref{Diffuse}(a)) and the introduction of third n.n. correlations does not improve the fit significantly. A high temperature expansion allows us to associate the ratio of correlations to the ratio of the corresponding exchange interactions~\cite{white1983,Hohlwein2003} so that we may infer that $J_2/J_1\approx-2.2(2)$. Even if some of the $J_1$ bond interactions are frustrated, this resulting fitted ratio of exchange parameters stabilize a $\mathbf{k}=(\frac{1}{2}\frac{1}{2}\frac{1}{2})$ order, which is driven by the dominant AFM $J_2$ interactions~\cite{Bartel1972,SUN2018176,Balla2020}.

Upon cooling, the elastic magnetic scattering gets stronger ($T=5.5{~\rm K}\approx T_N$ in Fig.~\ref{Diffuse}(b)) and eventually forms magnetic Bragg peaks ($T=1.6{~\rm K} << T_N$ in Fig.~\ref{Diffuse}(c)) indicating long range magnetic order. To quantify the temperature dependence of the diffuse and Bragg scattering, as shown in Fig.~\ref{Diffuse}(e), we fitted the integrated intensity obtained from one-dimensional (HHH) scans acquired through the magnetic Bragg peak at $\mathbf{Q}=(\frac{1}{2}\frac{1}{2}\frac{1}{2})$. Each scan was fit to the sum of a Gaussian function and a Lorentzian function to describe the long and short range components of the spin correlations, and a linear background (needed to describe the temperature independent nuclear and temperature-dependent magnetic incoherent elastic scattering). The fits included as dashed curves in Fig.~\ref{Diffuse}(e) provide a good account of the data.

The temperature dependence of the integrated intensity of both the Bragg and the diffuse components of the scattering are reported in Fig.~\ref{Diffuse}(f). The integrated intensity of the diffuse scattering (red markers) is peaked at $T_N$ where the appearance of Bragg scattering (blue markers) reveals the onset of long range order and translation symmetry breaking. The temperature variation of the correlation length $\xi$, as inferred from the Lorentzian after correcting for resolution effects, is reported in the inset of Fig.~\ref{Diffuse}(f). As expected, $\xi$ increases dramatically at $T_N$.

\subsection{Structural distortion}
\label{Xray}

A previous X-ray diffraction study revealed that a tetragonal distortion accompanies magnetic ordering in HoBi~\cite{hulliger1984low}. We confirmed the occurrence of this distortion in HoBi with a four-circle X-ray diffractometer experiment. The $\theta$-2$\theta$ scans of various nuclear Bragg peaks were collected above and below $T_N$ with a base temperature of 5~K. Consistent with previous work~\cite{hulliger1984low}, we observed a splitting of the $(H00)$, $(0K0)$, and $(00L)$ nuclear Bragg peaks whereas the $(HHH)$ Bragg peaks do not split. This indicates a tetragonal distortion and specifically precludes a rhombohedral distortion.  

The temperature dependence of a longitudinal $\theta$-2$\theta$ scan through the $\mathbf{Q}=(006)$ peak is plotted in Fig.~\ref{LowT}(a). This is an unfiltered copper source with  $K_{\alpha{_1}}$ and $K_{\alpha{_2}}$ radiation. Both components yield a split (006) peak below $T_N$. The distortion was quantified by fitting the $\theta$-2$\theta$ scans to Lorentzian functions while constraining the ratio of the $K_{\alpha{_1}}$ / $K_{\alpha{_2t}}$ integrated intensity to be temperature independent and set by its fitted value obtained at high temperatures. Examples of these fits are included in Fig.~\ref{LowT}(a). The temperature dependent lattice parameters inferred from this analysis are shown in Fig.~\ref{LowT}(b). The order parameter-like temperature dependence  is similar for both warming and cooling with no hysteresis detected down to the 100~mK temperature scale. For comparison the hysteresis detected through heat capacity measurements was 13~mK (Fig.~\ref{CpHoBi}). A single (006) Bragg peak with a lattice parameter of 6.2095(1)~$\angstrom$ above $T_N$, splits into two peaks with lattice parameters  6.2143(1)~$\angstrom$ and 6.2075(1)~$\angstrom$ below $T_N$. Assuming an approximately volume conserving phase transition implies that the lattice parameter that changes most is the $c$-axis. This indicates the structural unit cell elongates along the $c$-axis in the AFM state with $c/a$~=~1.0011(1) at 5~K. We note that an orthorhombic distortion with the $a$ and $b$ axis differing by less than 0.002~$\angstrom$ is not excluded by these data. 

A possible space group for HoBi below $T_N$ is the maximal tetragonal subgroup of the paramagnetic space group $Fm\overline{3}m$, which is $I4/mmm$. The structural parameters in the tetragonal phase are  $a_T=b_T=6.2075(1)/\sqrt{2}\angstrom$ and $c_T=6.2143(1)~\angstrom$ where the $\mathbf{a}_T$ and $\mathbf{b}_T$ axes are rotated by 45$\degree$ relative to the $\mathbf{a}$ and $\mathbf{b}$ axes of the paramagnetic simple cubic cell. In this space group Ho$^{3+}$ ions occupy a single $2a$ Wyckoff site and the magnetic ordering vector is $\mathbf{k}=(\frac{3}{2}0\frac{3}{2})$. While we must use the tetragonal space group below $T_N$, we continue to use the cubic unit cell to index wave vector transfer in the neutron scattering experiments, which do not resolve the multi-domain tetragonal distortion.

\subsection{Spin structure}

As described in the previous sections, the magnetic order has a characteristic wavevector $\mathbf{k}=(\frac{1}{2}\frac{1}{2}\frac{1}{2})$. In addition to the corresponding low $T$ magnetic Bragg peaks, the intensities of all nuclear Bragg peaks are observed to increase below $T_N$. The increase of intensity is approximately proportional to the intensity in the paramagnetic phase, which indicates it arises from secondary extinction release~\cite{bacon1948secondary}. To check this hypothesis, we performed polarized neutron diffraction on the (002) and (220) Bragg peaks below $T_N$ and found them to be exclusively nuclear in origin. 

We note that weak $\mathbf{k}=(001)$ Bragg peaks also onset at $T_N$. Examples of these peaks include the (001) and (111) Bragg peaks (see Fig.~\ref{Diffuse}(c)), which are forbidden within the $Fm\overline{3}m$ space group. These Bragg peaks are attributed to multiple magnetic scattering as their presence depends on both the employed incident neutron wavelength and the scattering plane, and they are absent in powder neutron diffraction measurements~\cite{FISCHER1985551}. The multiple scattering processes involve magnetic $\mathbf{k}=(\frac{1}{2}\frac{1}{2}\frac{1}{2})$ Bragg reflections so they occur only for $T<T_N$.

\begin{figure}[tbp]
\linespread{1}
\par
\includegraphics[width=3.4in]{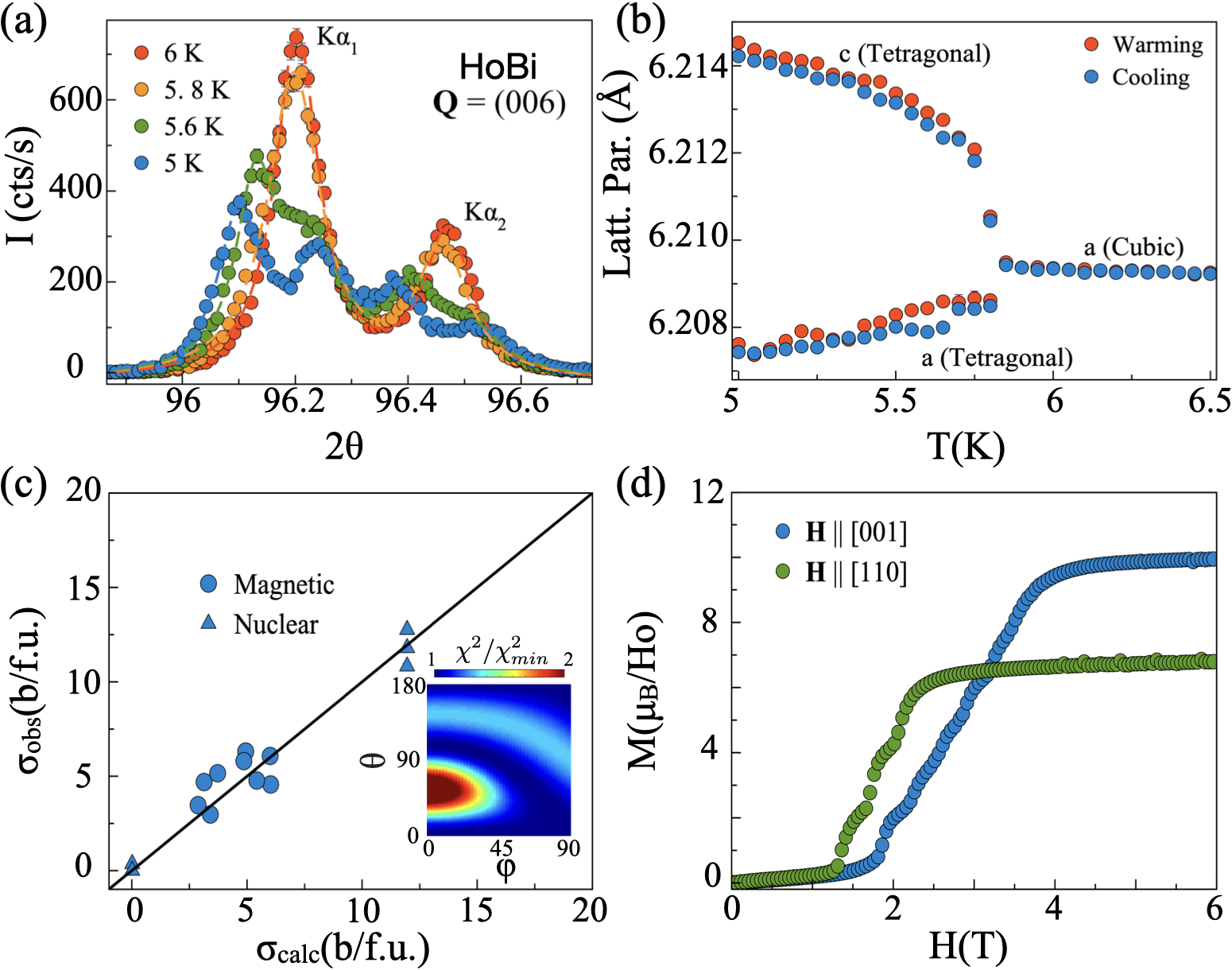}
\par
\caption{A series of $\theta$-2$\theta$ X-ray diffraction scans through the $\mathbf{Q}$~=~(006) Bragg peak. The inferred temperature dependence of the lattice parameters is shown in panel (b). The neutron magnetic and nuclear refinement of HoBi are presented in (c) where the observed cross-sections for various Bragg peaks are plotted as a function of the calculated cross-sections. The inset in (c) reports the variation of the $\chi^2$ goodness of fit for the magnetic refinement of HoBi assuming a multi-domain $\mathbf{k}~=~(\frac{1}{2}\frac{1}{2}\frac{1}{2})$ spin structure with an easy axis defined by spherical coordinates $\theta$ and $\phi$ ($\phi$~=~0 corresponds to the [110] direction). Panel (d) shows the low-temperature magnetization versus field for fields applied parallel to the [001] and [110] directions. The data show that [001] is the easy axis.}
\label{LowT}
\end{figure}

Referring to $fcc$ close packing, the AFM $\mathbf{k}~=~(\frac{1}{2}\frac{1}{2}\frac{1}{2})$ spin structure can be described as an AFM stacking of FM triangular lattices. As the magnetic order and structural distortion in HoBi occur in a single 1st order phase transition, the direction of the spins in each FM sheet is not constrained by the usual Landau argument for second order phase transitions. To determine the local spin orientation of the Ho$^{3+}$ ions, we collected 18 rocking scans at different magnetic Bragg positions for a sample presumed to be in an unbiased multi-domain state. The data were compared to a cubic domain average of the calculated magnetic Bragg diffraction for a general spin orientation within one domain given by spherical angles $\theta, \phi$ and $\mathbf{k}=(\frac{1}{2}\frac{1}{2}\frac{1}{2})$. Here $\theta=0$ corresponds to the tetragonal $c$-direction and $\theta=\pi/2$ and $\phi=0$ corresponds to the [110] direction. Minimizing with respect to the moment size at each point, the $\chi^2$ measure of fit quality is shown versus $\theta$ and $\phi$ in the inset panel of Fig.~\ref{LowT}(c). The manifold of states represented by the red arrows in Fig.~\ref{NucStruc} are indistinguishable by neutron diffraction. This degeneracy arises because the magnetic diffraction intensity for a multi-domain sample only depends on the smallest angle between the spin and a $\langle 111 \rangle$ axis. From our refinement, we find this angle is 47(10)$\degree$. This is experimentally indistinguishable from the angle between [001] and [111], which is 55$\degree$. This means the magnetic diffraction data are consistent with spins pointing along the [001] directions, but also with many other directions including close to the [110] direction. 

Fortunately the spin anisotropy of the Ho$^{3+}$ ions can be deduced from other pieces of information. First, the low-temperature magnetization of HoBi shown in Fig.~\ref{LowT}(d) reveals the saturation magnetization is larger for fields along the [001] direction than along [110]. Second, the structural distortion also occurs along the [001] direction. Both of these measurements are consistent with spins oriented along the tetragonal $c_T$-axis in the AFM ordered state. Additionally, in Sec.~\ref{lowE} we show that a [001] easy axis anisotropy is needed to accurately model the inelastic neutron scattering spectrum below $T_N$. We thus conclude the spins in the AFM type II order of HoBi are oriented along the $\mathbf{c}_T$ direction, which is the direction of the structural elongation. The comparison between measured and calculated magnetic Bragg intensities is shown in Fig.~\ref{LowT}(c). The corresponding spin structure is shown in Fig.~\ref{NucStruc}. An ordered moment of 10.3(6)~$\mu_B$ was determined, which is experimentally indistinguishable from the $gJ\mu_B=\frac{5}{4}\cdot 8~\mu_B=10~\mu_B$ saturation magnetization of Ho$^{3+}$. 

\subsection{Crystal electrical field interaction}

For Ho$^{3+}$ ions, the $J=8$ spin-orbit ground state manifold is (2$J$+1)~=~17 fold degenerate under full rotation symmetry. This degeneracy is, however, lifted by the symmetry breaking crystal electric fields (CEF). Using the Stevens operator formalism, the CEF Hamiltonian appropriate for Ho$^{3+}$ in the high-temperature cubic phase of HoBi can be expressed as follows:
\begin{equation}
\hat{H}_{cef}^{cubic} = B_4(\hat{O}_4^0+5\hat{O}_4^4) + B_6(\hat{O}_6^0-21\hat{O}_6^4).
\end{equation}
Here $\hat{O}_n^m$ are Stevens operators~\cite{stevens1952matrix} that can be written in terms of the spin-orbital angular momentum operators $\hat{J}_+$, $\hat{J}_-$ and $\hat{J}_z$ where $\hat{\bf z}\parallel {\bf c}$. The CEF parameters $B_n$ are scalars of dimension energy that dictate the strength of the different CEF terms and can be determined by fitting spectroscopic or thermo-magnetic data sensitive to the crystal field level scheme. $B_n$ can also be estimated through the point-charge model~\cite{hutchings1964point}. 

Following  Hutching's formalism~\cite{hutchings1964point} the point charge model yields
\begin{equation}
B_4~=~\frac{7 |e| |q_{Bi}| \beta_J \langle r^4 \rangle}{64 \pi \epsilon_0 d_{Bi}^5}
\label{eqB4}
\end{equation}
and
\begin{equation}
B_6~=~\frac{3 |e| |q_{Bi}| \gamma_J \langle r^6 \rangle}{256 \pi \epsilon_0 d_{Bi}^7}.
\label{eqB6}
\end{equation}
Here $e$ is the electron charge, $q_{Bi}$ is the charge of the Bi ligand and $\epsilon_0$ is the vacuum permitivity. $\beta_J$ and $\gamma_J$ are reduced matrix elements calculated in ref~\cite{stevens1952matrix} whereas the radial integrals for the $4f$ state $\langle r^n \rangle$ are tabulated in ref~\cite{freeman1962theoretical}. We used $q_{Bi}~=~-3e$ and the distance between a holmium ion and its first n.n. bismuth ion $d_{Bi}=a/2=6.2093(1)/2$~\AA. Introducing these values in Eqs.~\ref{eqB4} and \ref{eqB6} we obtain $B_4=-2.2709(2)\times 10^{-4}$~meV and $B_6=-1.0468(1) \times 10^{-7}$~meV.  

\begin{figure}[!ht]
\linespread{1}
\par
\includegraphics[width=3.4in]{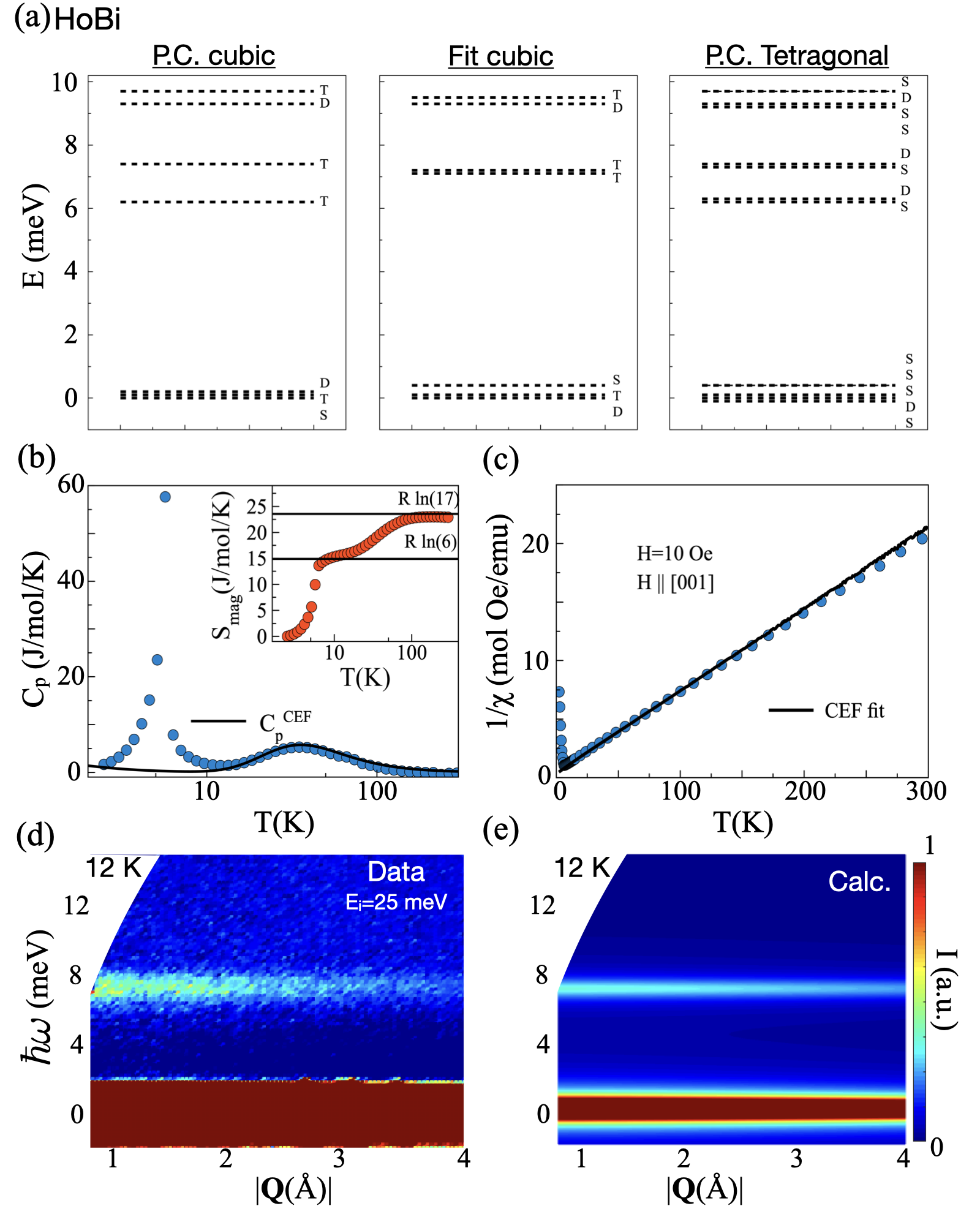}
\par
\caption{Determination of the crystal electric field (CEF) level scheme for the $J$=8 Ho$^{3+}$ ion in HoBi. (a) shows the results of a point charge (PC) calculation for the cubic and tetragonal phases. The cubic CEF scheme may be compared to the level scheme for the fitted CEF Hamiltonian of HoBi. Panel (b) and (c) respectively show the temperature dependence of the magnetic heat capacity ($C_p$) and the inverse magnetic susceptibility of HoBi compared to corresponding properties based on the fitted CEF Hamiltonian. The magnetic entropy obtained from integrating the $C_p$ of HoBi is shown in the inset of (b). The measured (d) and calculated (e) inelastic neutron scattering spectra of HoBi are shown for $T~=~12~K$. The neutron inelastic scattering data were acquired using a 25~meV incident neutron beam.}
\label{CEF} 
\end{figure}

The corresponding CEF level scheme for Ho$^{3+}$ in the cubic phase of HoBi is shown in Fig.~\ref{CEF}(a). The Ho$^{3+}$ $J-$multiplet is split into 4 triplets, 2 doublets, and 1 singlet that form three groups. Group I includes one doublet, one triplet, and one singlet between 0 and 0.2~meV. Group II is formed by two triplets between 6~meV and 7~meV, and group III consists of a doublet and a triplet between 9~meV and 10~meV.  

The CEF Hamiltonian estimated from our point-charge calculation can reproduce the temperature dependence of the magnetic heat capacity $C_p$ (Fig.~\ref{CEF}(b)) and magnetic susceptibility $\chi$ (Fig.~\ref{CEF}(c)). Obtained by integrating $C_p/T$, the temperature dependence of the entropy shown in the inset of Fig.~\ref{CEF}(b) is informative. A first entropy plateau near 10~K is associated with the sharp $C_p$ anomaly at the phase transition to long range magnetic order. The corresponding change in entropy of $\Delta S~=~R\ln{6}$ is that associated with the group I CEF states. The second plateau at $S~=~R\ln{17}$ is reached at room temperature and encompasses all of the entropy associated with the three groups of crystal field levels. 

For a more stringent test of the point charge model, we turn to inelastic neutron scattering. Fig.~\ref{CEF}(d) shows the 12~K inelastic neutron scattering spectrum with energy transfer ranging from 0 to 15~meV. At this temperature, the group II and III of CEF states are so scarcely populated that only CEF excitations originating from group I should be visible. No significant intrinsic broadening of the CEF excitations is observed and we note, also, that the experimental resolution is too coarse to resolve CEF levels within a group. The magnetic neutron scattering cross section associated with CEF transition from group I to II and from group I to III can be computed based on the point charge CEF Hamiltonian ($I_{mn}\propto~\sum_i|\bra{m} J_i \ket{n}|^2$). This calculation predicts the cross section for transitions from group I to group II is 250 times stronger than for transitions from group I to group III. The intensity of the transition from I to III is thus predicted to be too weak to be detected. This explains why Fig.~\ref{CEF}(d) shows just a single peak that we associate with transitions from group I to group II crystal field levels.

While the measured 7.2 meV gap between group I and group II CEF levels is just 0.4 meV off from the point charge prediction of 6.8 meV, we can improve our estimate of the CEF Hamiltonian by simultaneously fitting $B_4$ and $B_6$ for the best possible account of the neutron scattering spectra (Fig.~\ref{CEF}(d)), the specific heat data (Fig.~\ref{CEF}(b)), and the magnetic susceptibility data (Fig.~\ref{CEF}(c)). The best fit parameters thus obtained are $B_4~=~-2.24(1) \times  10^{-4}$~meV and $B_6~=~-2.4(1) \times 10^{-7}$~meV and with them the CEF Hamiltonian provides an excellent account of all single ion properties that we've measured, as shown in Fig.~\ref{CEF}. 

The CEF scheme obtained from our fit (Fig.~\ref{CEF}(a)) is remarkably similar to the point-charge calculation. Also a re-scaling of our CEF Hamiltonian for HoBi using Eq.~\ref{eqB4} and Eq.~\ref{eqB6} considering only the different ligand spacing successfully predicts the level scheme for HoN ref~\cite{furrer1976crystal}. This is in contrast with the praseodymium case where a pnictide ligand charge of $q=-2e$ is needed to bring the point charge model into agreement with experimental data~\cite{Birgeneau1973}. This indicates that holmium monopnictides are more ionic than praseodymium monopnictides. 

Finally, we estimated the effect of the tetragonal distortion on the CEF interaction in HoBi. We performed a point-charge calculation assuming that the first n.n. Ho-Bi bond is shorter along the $\mathbf{a}$ and $\mathbf{b}$ direction ($d_a$) as compared to the $\mathbf{c}$ direction ($d_c$). The calculated CEF Hamiltonian can be written as:

\begin{align}
        {\cal H}^{tet}_{cef} = \frac{|e||q_{Bi}|}{4\pi\epsilon_0} [\alpha_J \langle r^2 \rangle                 (\frac{1}{d_{c}^3}-\frac{1}{d_{a}^3})\hat{O}_2^0 + \\
        \nonumber
         \beta_J \langle r^4 \rangle ((\frac{1}{4 d_{c}^5}+\frac{3}{16d_{a}^5})\hat{O}_4^0 + \frac{35}{16d_{a}^5}\hat{O}_4^4) + \\ 
         \nonumber
         \gamma_J \langle r^6 \rangle ((\frac{1}{8d_c^7}-\frac{5}{64d_a^7})\hat{O}_6^0-\frac{63}{64d_a^7}\hat{O}_6^4)].
\end{align}

The corresponding level scheme is shown in Fig.~\ref{CEF}(a). For this calculation, we used the lattice parameters determined in our high-resolution X-ray scattering experiment. The degeneracy of all the triplets and doublets associated with cubic symmetry is lifted. This results in four doublets and nine singlets and a significant broadening of each of the three groups of crystal field levels.

\subsection{Low energy spin dynamics}
\label{lowE}

\begin{figure}[tbp]
\linespread{1}
\par
\includegraphics[width=3.5in]{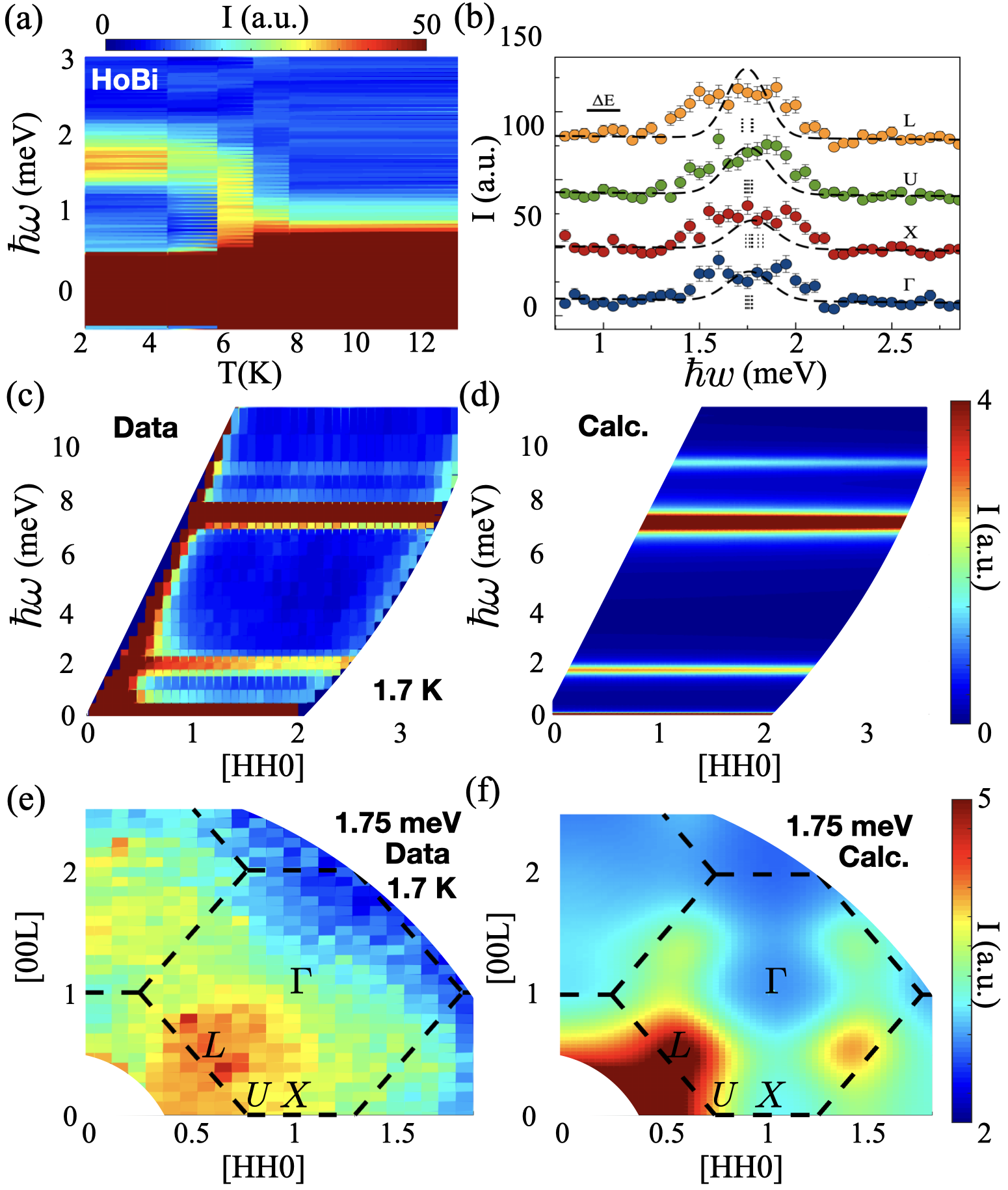}
\par
\caption{The temperature dependence of the low energy inelastic neutron spectrum of HoBi at $\mathbf{Q}=(\frac{1}{2}\frac{1}{2}\frac{1}{2})$ is shown in panel (a). The spectrum of neutron scattering at some high symmetry positions within the first Brillouin zone of HoBi are shown in (b).The horizontal black bar indicates the FWHM energy resolution of the spectrometer while the black dashed lines show the predicted spectrum based on the spin Hamiltonian presented in this work. The energies associated with each exciton are indicated by vertical black dashed lines. The observed and calculated inelastic neutron scattering spectrum up to 11.5~meV are respectively plotted in (c) and (d) for momentum transfer $\mathbf{Q}$ along the [HH0] direction. The observed and calculated momentum dependence of the 1.75~meV exciton scattering intensity is shown in (e) and (f). The energy integration for panel (e) is $\pm$0.25~meV.}
\label{SpinDyn} 
\end{figure}

We now turn our attention to the collective physics of HoBi, which we explore using inelastic magnetic neutron scattering. Fig.~\ref{SpinDyn}(a) shows the temperature dependence of the inelastic scattering for $\mathbf{Q}=(\frac{1}{2}\frac{1}{2}\frac{1}{2})$. Just above $T_N$, the scattering is quasi-elastic with a physical (resolution corrected) FWHM of 0.30(5)~meV. No inelastic intensity is observed up to 2~meV. This is consistent with the  CEF energy scheme shown in Fig.~\ref{CEF}(c). Below $T_N$, the  quasi-elastic scattering splits into an elastic and an inelastic component.

To probe any dispersion of the low energy spin excitations, we acquired low energy spectra at momentum transfer $\mathbf{Q}$ corresponding to high symmetry points in the Brillouin zone. Fig.~\ref{SpinDyn}(b) shows the spectrum consists of a peak that is broader than the experimental resolution (FWHM indicated by horizontal bar) and that shifts by less than the peak width between the different values of $\mathbf{Q}$. A gaussian fit finds the peak centered at 1.7(2)~meV with a FWHM of 0.48(4)~meV that exceeds the instrumental resolution (FWHM of 0.22~meV). The limited resolution and statistical accuracy of the data does not rule out the possibility of multiple dispersive components within the approximately Gaussian envelope of the peak.

We also examined the higher energy excitations for $T<T_N$ by acquiring momentum resolved inelastic scattering data up to 11.5~meV. A representative slice through the data is displayed as a color image versus $\mathbf{Q}$ along the (HH0) direction and energy transfer in Fig.~\ref{SpinDyn}(c). No dispersion is resolved. The data are similar to the high-temperature plot of intensity versus $|Q|$ and $\hbar\omega$ in Fig.~\ref{CEF}(d) though with  additional inelastic features at 9.0(3)~meV and 1.7(2)~meV.

Fig.~\ref{SpinDyn}(e) shows the momentum dependence of the integrated intensity of the 1.7 meV mode throughout the $(HHL)$ zone. The $\mathbf{Q}$ dependence of the intensity is subtle albeit peaked at the magnetic $(\frac{1}{2}\frac{1}{2}\frac{1}{2})$ zone center and smoothly decreases with $|Q|$ in accordance with the Ho$^{3+}$ magnetic form factor~\cite{Blume1962}. We note that the 1.7 meV gap is about an order of magnitude greater than the predicted CEF gap arising from the tetragonal distortion. This indicates the phase transition is driven by the magnetic interactions, which we model below.

\section{Modeling spin dynamics of spin-orbital excitons}

The low-temperature excitations in HoBi are similar to other rare-earth metallic compounds where exchange interactions are strong enough to mix crystal field levels~\cite{rainford1971magnetic,Turberfield1971,buyers1971excitations}. Because  components that are longitudinal with respect to the ordered moment are involved, these are not conventional transverse spin wave excitations. They may be described as crystal field excitations that can propagate through the lattice due to inter-site interactions. We shall adopt the practice of calling these ``crystal field exciton'' or simply ``exciton''~\cite{sarte2019,sarte2020van,Yuan2020}.

A common theoretical approach to describing excitons in rare-earth magnets is to use a pseudo-boson theory where the exciton creation operator is a linear combination of single-ion operators~\cite{buyers1971excitations,holden1974magnetic,bak1975magnetic}. In this theory, the $\mathbf{Q}={\bf 0}$ single-ion operators are obtained by diagonalizing the mean-field spin Hamiltonian and the dispersion at finite $\mathbf{Q}$ is produced by the exchange terms. We use this pseudo-boson theory to describe the magnetic excitation spectrum of HoBi below $T_N$. The Hamiltonian ${\cal H}_s$ includes the single-ion tetragonal crystal field terms and isotropic exchange interactions. ${\cal H}_s$ is decomposed into a mean-field term (${\cal H}_{0,k}$) and an interacting part (${\cal H}_{int}$) so ${\cal H}_s=\sum_k{\cal H}_{0,k}+{\cal H}_{int}$ where: 
\begin{align}
        {\cal H}_{0,k} = {\cal H}^{tet}_{cef,k} + (-1)^k H_z J_{jz}^k
\end{align}

and 

\begin{align}
        {\cal H}_{int} = \sum_{j,j',k,k'} {\cal J}^{k,k'}_{j,j'}  \mathbf{J}^k_j\cdot\mathbf{J}^{k'}_{j'} - \sum_{j,k}(-1)^k H_z J_{jz}^k.
\end{align}

Here $j$ indexes the unit cell while $k=1,2$ specifies the anti-parallel sub-lattices of the AFM order (Fig.~\ref{NucStruc}). We define $H_z=2\sum_{r}Z_r {\cal J}_r\langle J_z\rangle$ where ${\cal J}_r$ and $Z_r$ are respectively the exchange constant and coordination number associated with the $r^{th}$ neighbor. $\langle J_z\rangle$ is the thermal average of $J_z$ on each site, which we found to be $\langle J_z\rangle=8$ in our diffraction and CEF analysis. By definition, ${\cal H}_{int}$ carries no mean value and so can be written in terms of creation ($\hat{a}_{n,k}^\dagger$~=~$\ket{n,k}\bra{0,k}$) and annihilation ($\hat{a}_{n,k}$~=~$\ket{0,k}\bra{n,k}$) operators that connect the ground state $\ket{0,k}$ and the excited eigenstates $\ket{n,k}$ of $\hat{H}_{0,k}$. In this case, $\hat{H}_{0,k}~=~\sum_n E_n\hat{a}_{n,k}^\dagger\hat{a}_{n,k}$ where $E_{n,k}$ is the eigenvalue of the $\ket{n,k}$ eigenstate of $\hat{H}_{o,k}$. After writing $\hat{H_s}$ in terms of these operators and Fourier transforming it, we obtain:
\begin{align}
        \hat{H}_s = \frac{1}{2}\sum_{\mathbf{Q}} \bigg\{ \hat{a}^\dagger(\mathbf{Q}) A(\mathbf{Q})\hat{a}(\mathbf{Q}) + \hat{a}^\dagger(-\mathbf{Q}) A (-\mathbf{Q})\hat{a}(-\mathbf{Q})\\
        \nonumber
        + \hat{a}^\dagger(\mathbf{Q}) B(\mathbf{Q})\hat{a}^\dagger(-\mathbf{Q}) + \hat{a}(-\mathbf{Q})B(\mathbf{Q})\hat{a}(-\mathbf{Q}) \bigg\}
\end{align}

with $\hat{A}$~=~$\hat{\Delta}$ + 2$\hat{h}_{zz}$ + $\hat{h}_{+-}$ + $\hat{h}_{-+}$ and $\hat{B}$~=~2$\hat{h}_{zz}$ + $\hat{h}_{++}$ + $\hat{h}_{--}$ where $\hat{\Delta}$~=~$E_{n,k}\delta_{k,k'}\delta_{n,n'}$ and $\hat{h}_{\alpha\beta}(k,k',n,n',\mathbf{Q})$~=~$J(\mathbf{Q})\bra{k,n} \hat{J}_\alpha \ket{0,k} \bra{k',0} \hat{J}_\beta \ket{n',k'}$.

The procedure to compute the spin dynamics first consist of diagonalizing $\hat{H}_{0,k}$ to obtain the eigenvalues E$_{n,k}$ and eigenvectors $\ket{n,k}$ for $\mathbf{Q}~=~{\bf 0}$. At finite $\mathbf{Q}$, the matrix $\hat{H}_s$~=~$\begin{pmatrix}
  \hat{A} & \hat{B}\\ 
  -\hat{B} & -\hat{A}
\end{pmatrix}$ is then computed and diagonalized to obtain the perturbed energies ($E_{\tilde{n}}(\mathbf{Q})$) and eigenstates $\ket{\tilde{n}(\mathbf{Q})}$ for each exciton. We consider all the excited CEF states belonging to the (2$J$+1) spin-orbit manifold of HoBi so there are 32 creation and annihlation operators for each of the 2 Ho$^{3+}$ spins within the magnetic unit cell. This give a Hilbert space of 64 states for $\hat{H}_s$. The associated inelastic magnetic neutron scattering cross-section for a single magnetic domain is then~\cite{buyers1971excitations,holden1974magnetic}:
\begin{eqnarray}
\cfrac{d^2 \sigma}{d E d \Omega}&=&N (\gamma r{_0})^2\frac{k_f}{k_i}|\frac{g}{2}f(\mathbf{Q})|^2\\
&&\times { \sum_{\tilde{n},\mathbf{q},\mathbf{\tau_m}} |\bra{\tilde{n}(\mathbf{q})}\hat{J}_{\mathbf{Q}}\ket{GS}|^2\delta(E-E_{\tilde{n}}(\mathbf{q})) \Delta(\mathbf{Q}-\mathbf{q}-\mathbf{\tau_m})} \nonumber
\label{eqInt}
\end{eqnarray}
Here $N$ is the number of primitive magnetic unit cells, $\gamma$~=~-1.91 is the gyromagnetic ratio of the neutron, $r_0~=~2.818 \times 10^{-15}$~m is the classical electron radius, $\mathbf{\tau_m}$ is the magnetic zone center, $\mathbf{q}$ is the reduced momentum transfer within the first magnetic Brillouin zone, while $k_f$ and $k_i$ respectively are the scattered and incoming neutron wave vector. The measured spectrum is subject to the finite resolution of the instrument which we account for by replacing the delta functions by a united normalized Gaussian functions with the $\mathbf{Q}$-integrated energy resolution width. The final calculated spectrum was averaged over all possible magnetic domains.

\section{Microscopic Spin Hamiltonian for Holmium Bismuth}

We determined the microscopic parameters of $\hat{H}_s$ for HoBi by fitting the $\mathbf{Q}~=~{\bf 0}$ spectrum consisting of three excitons at $E_1$~=~1.7(2)~meV, $E_2$~=~7.4(2)~meV and $E_3$~=~9.0(3)~meV with relative intensities $I_2/I_1$~=~5.5(3) and $I_2/I_3$~=~37(7). Employing the ratio $\|J_2/J_1\|$~=~2.17 obtained by analyzing the magnetic diffuse scattering (section \ref{diffusetext}) leaves just one free parameter. The tetragonal CEF Hamiltonian has six free parameters that were initially estimated from the point-charge model. To reproduce the exact energies of the excitons at $E_2$ and $E_3$, we allowed the CEF parameters to relax away from their point-charge values which results in many combinations of parameters consistent with the data. We estimated the exchange constants by varying the CEF parameters away from their point-charge calculation values and keeping all solutions that have a $\chi^2$ within 20$\%$ (1/N$_{obs}$) of the global minimum. The exchange parameters refined to $J_1~=-~1.4(2)~\mu eV$ and $J_2~=~~3.0(5)~\mu eV$. A mean-field critical temperature of 20(7)~K is obtained from these parameters. For comparison, the actual ordering temperature is only $T_N~=~5.72(1)~$K. We hypothesize that fluctuations arising from competition between the ferromagnetic $J_1$ and the antiferromagnetic $J_2$ interactions lead to the reduced critical temperature. 

The right column of Fig.~\ref{SpinDyn} compares the optimized model for a multi-domain sample to the experimental data. Fig.~\ref{SpinDyn}(d) shows the full intensity versus $\hbar\omega$ and $\mathbf{Q}\parallel(HH0)$ for comparison with Fig.~\ref{SpinDyn}(c). The position and relative intensity of the three modes are well reproduced. Looking more closely at the 1.75~meV mode, Fig.~\ref{SpinDyn}(b) compares the intensity versus energy transfer at select high symmetry points in the Brillouin zone. The vertical dashed lines show that multiple excitons contribute at each $\mathbf{Q}$. This is generally consistent with the featured spectrum observed though there is more broadening/dispersion observed than reproduced by the model. Inclusion of anisotropic or longer range interactions might be needed to remedy this discrepancy though data with higher energy resolution is needed to justify the greater model complexity. Fig.~\ref{SpinDyn}(f) shows the calculated $\mathbf{Q}$-dependent integrated intensity of the 1.75~meV mode. The dominant features of the experimental result in Fig.~\ref{SpinDyn}(e) are reproduced, including mainly the increase of scattered intensity at the magnetic zone centers. We note the presence of phonon scattering near $\mathbf{Q}~=~(002)$ that may account for the discrepancy between the calculation and the experimental data at that momentum point. 

\section{Discussion and Conclusion}
In this manuscript, we have characterized an antiferromagnetic order and the associated crystal field excitons that develop below T$_N~=~5.72(1)~K$ in the rare-earth monopnictide HoBi. This magnetic state is driven by strong 2nd n.n. antiferromagnetic and weaker 1st n.n. ferromagnetic interactions, which we quantified via modeling of the diffuse paramagnetic and low temperature inelastic neutron scattering. The excitation spectrum is sensitive to the local orientation of the Ho$^{3+}$ ordered spins, which allowed us to establish the Ising nature of the antiferromagnetic order in HoBi that cannot be deduced from neutron diffraction of a multi-domain sample. We used X-ray diffraction to provide evidence for a tetragonal structural distortion that accompanies magnetic ordering. Our CEF analysis and modelling of inelastic scattering data indicates the elongated $c$-axis coincides with the easy magnetic axis within a domain.\\

The magnetic excitations that we have documented here surely have significant impacts on the magneto-transport properties of HoBi~\cite{yang2018interplay}. For example, we found strong quasi-elastic neutron scattering in the paramagnetic state. The associated short range correlated spin fluctuations, which may be accompanied by short range tetragonal lattice distortions too given the non-Kramers nature of the Ho$^{3+}$, are expected to enhance the electrical resistivity above $T_N$. Below $T_N$, these gapless fluctuations are replaced by a coherent exciton at 1.7(2)~meV and correspondingly the electrical resistivity is reduced by an order of magnitude upon cooling below $T_N$~\cite{yang2018interplay}. The field-dependence of spin-orbital excitons may be responsible for various features observed in the magnetoresistance of HoBi and more broadly in the rare-earth monopnictides~\cite{liang2018extreme,wu2017large,ye2018extreme,wang2018extremely,wang2018unusual,lyu2019magnetization,wu2019multiple}.\\

\section{acknowledgements}
This work was supported as part of the Institute for Quantum Matter, an Energy Frontier Research Center funded by the U.S. Department of Energy, Office of Science, Basic Energy Sciences Under Award No.DE-SC0019331. CB was further supported by the Gordon and Betty Moore foundation EPIQS program under GBMF9456. The work at Boston College was supported by the U.S. Department of Energy, Office of Basic Energy Sciences, Division of Physical Behavior of Materials under Award DE-SC0023124. This work was supported in part by the Natural Sciences and Engineering Research Council of Canada (NSERC). We acknowledge the support of the National Institute of Standards and Technology, U.S. Department of Commerce. Access to MACS was provided by the Center for High Resolution Neutron Scattering, a partnership between the National Institute of Standards and Technology and the National Science Foundation under Agreement No. DMR-1508249. The identification of any commercial product or trade name does not imply endorsement or recommendation by the National Institute of Standards and Technology. A portion of this research used resources at the High Flux Isotope Reactor, a DOE Office of Science User Facility operated by the Oak Ridge National Laboratory.

\bibliography{HoBiBib}

\end{document}